\documentstyle[aps,12pt,psfig]{revtex}
\begin{document}
\tightenlines
\title{Inflation from susy quantum cosmology}
\author{W. Guzm\'an, J. Socorro, V. I. Tkach and J. Torres}
\address{Instituto de F\'{\i}sica de la Universidad de Guanajuato,\\
 A.P. E-143, C.P. 37150, Le\'on, Guanajuato, M\'exico}

\maketitle
\widetext

\begin{abstract}
We propose a realization of inverted hybrid inflation scenario in the 
context of $n=2$ supersymmetric quantum cosmology. The spectrum of density 
fluctuations is calculated in the de Sitter regimen as a function of the 
gravitino and the Planck mass, and  explicit forms for the wave function 
of the universe are found in the WKB regimen for a FRW closed and flat 
universes.

\end{abstract}

\narrowtext

\section{Introduction}

Inflation is a theoretical attractive idea for solving many classical problems 
of the Standard Big Bang Cosmology. Recently, observations have confirmed its 
predictions of a flat universe with a nearly scale invariant perturbations 
spectrum \cite{1,2}. 

On the other hand, despite its many successes, there are 
still not completely natural  inflation models known in particle physics. 
Inflation generally requires small parameters in particle theory to provide a 
flat potential, needed for sufficient inflation and for the correct density 
fluctuations \cite{1,3}. Probably the most attractive models of inflation are 
the hybrid inflation \cite{c,4,5,6}, extranatural inflation \cite{7,8} 
and inverted hybrid inflation \cite{a,b} models. In the hybrid inflation 
scenario proposed by Linde \cite{c,4}, the slowly rolling 
inflaton field $\phi$ is not the responsible for most of the energy density. 
That role is played by another field $\psi$, which is held in place by its 
interaction with the inflaton field until the 
latter falls below a critical value, thus, $\psi$ is 
destabilized, rolling to its true vacuum, and inflation ends.
The hybrid inflation models are based on particle physics motivation 
such as the supersymmetry, supergravity, and superstrings. In the context of 
the last two of these, we do not expect inflation to be possible for field 
values exceeding a Planck mass, regardless
of whether the potential energy there is larger than the Planck energy, 
because supergravity corrections tend to generate a steep potential 
that is unable to sustain inflation. Given the success of hybrid inflation, it 
was subsequently suggested that the hybrid mechanism could be adapted to
create an inverted model in which the inflaton field $\phi$ has a negative
mass aquared and rolls away from the origin, predicting a spectral index
which can be significantly below 1 in constrast to virtually all other
hybrid inflation models.

Supersymmetry may play an important role for inflation. Many models have been 
proposed describing the inflationary phase transition in globally 
supersymmetric theories \cite{3,5,9,10,11,12} and locally supersymmetric 
theories \cite{5,13,14,15,16}. This models have been analized 
without and with supergravity corrections. Supergravity corrections spoil the 
flatness of the inflaton potential because supersymmetry is broken during 
inflation.

The study of supersymmetric minisuperspace models has lead to important and 
interesting results \cite{17}. In works \cite{18,19,20} a new approach have 
been proposed to the study of 
supersymmetric quantum cosmology. The main idea is to extend the group 
of local time 
reparametrization of the cosmological models to the local conformal time 
supersymmetry, which is a subgroup of the four dimensional spacetime 
supersymmetry. This scheme allowed us to formulate, in the superfield 
representation, the supersymmetric action to study homogeneous models. 
The Grassmann superpartners of 
the scale factor and the homogeneous scalar fields at the quantum level, 
are elements 
of the Clifford algebra.  In this level, these models are  a specific 
supersymmetric quantum mechanics models  with spontaneous breaking of 
supersymmetry when the vacuum energy is zero.
In this paper we consider the inverted hybrid inflation models in the 
context of local conformal $n=2$ supersymmetry and we find some simple 
solutions (WKB solutions) to the Wheeler-De Witt equation.

The plan of this paper is as follow. In Sec. II we introduce the n=2 local
conformal supersymmetry formulation of the FRW model interacting with a set
of spatially homogeneous real scalar matter superfields. In Sec. III
we introduce the WKB procedure to obtain classical and quantum solutions,
and also, we shall analize the potential proposed in our model, in the 
case of the inflationary potential where this is modified by the presence of 
the local supersymmetry breaking sector (via the scalar $\varphi$). Also,
we consider the semiclassical solutions to the WDW equation for 
the inflationary phase (not zero potential and zero superpotential),  and 
the potential is zero but the  superpotential is 
not zero, both cases are relationed with the two minimums that have the
potential, local  and global minimums, respectively. Besides, the density 
fluctuation produced by the inflaton field
$\phi$, is shown. The Sec. IV is devoted to conclusions.

\section{ Supersymmetric Lagrangian and susy breaking}

The most general superfield action for a homogeneous scalar supermultiplet 
interacting with the scale factor in the supersymmetric $n=2$ FRW model has 
the form
\begin{equation}
S=\int \left[ 6\left(-\frac{1}{2\kappa^2}\frac{{\cal R}}{\cal N}
D_{\bar{\eta}}{\cal R}
D_{\eta}{\cal R}+\frac{\sqrt{k}}{2\kappa^2}{\cal R}^2\right)
+\frac{1}{2}\frac{{\cal R}^3}{\cal N}D_{\bar{\eta}}\Phi^i D_{\eta}\Phi^i 
-2{\cal R}^3g(\Phi ^i) \right]d\eta d\bar{\eta} dt,
\label{eqn1}
\end{equation}
where $k=0,1$ denotes flat and closed space and 
$\kappa^2=8\pi G_N=\frac{1}{M_p^2}$, $G_N$ is
the Newton's gravitational constant ($\hbar =c=1$). The units for the 
constants and fields  are the following: 
$[\kappa^2]=l^2,[{\cal N}]=l^0,[{\cal R}]=l^1,
[\Phi^i]=l^{-1},[g(\Phi^i)]=l^{-3}$, here $l$ correspond to units of 
length. 
$D_{\eta}=\frac{\partial}{\partial\eta}+
i\bar{\eta}\frac{\partial}{\partial t}$ and 
$D_{\bar{\eta}}=-\frac{\partial}{\partial\bar{\eta}}
-i\eta\frac{\partial}{\partial t}$, 
$i=1,2,3$, are the supercovariant derivatives of the superconformal 
supersymmetry $n=2$,
which have dimensions of  $[D_{\eta}]=[D_{\bar{\eta}}]=l^{-1/2}$.

Its series expansion for the one-dimensional gravity superfield 
${\cal N}(t,\eta ,\bar{\eta})$  is
\begin{equation}
{\cal N}=N(t)+i\eta \bar{\psi}'(t)+i\bar{\eta}\psi '(t)+\eta\bar{\eta}V'(t),
\label{eqn2}
\end{equation}
in which, $N(t)$ is the lapse function, and we have also introduced the 
reparametrization $\psi '(t)=N^{1/2}(t)\psi (t)$ and 
$V'(t)=N(t)V(t)+\bar{\psi}(t)\psi (t)$.

The Taylor series espansion for the superfield ${\cal R}(t,\eta ,\bar{\eta})$ 
has a similar form
\begin{equation}
{\cal R}=R(t)+i\eta \bar{\lambda}'(t)+i\bar{\eta}\lambda '(t)
+\eta\bar{\eta}B'(t),
\label{eqn3}
\end{equation}
where $\lambda '(t)=\kappa N^{1/2}(t)\lambda (t)$ and 
$B'(t)=\kappa N(t)B(t)+\frac{1}{2}\kappa 
(\bar{\psi}\lambda -\psi\bar{\lambda})$.

The real scalar matter superfields $\Phi^i$ may be written as
\begin{equation}
\Phi^i =\phi^i(t)+i\eta \bar{\chi '}^{i}(t)+i\bar{\eta}\chi '^{i}(t)
+\eta\bar{\eta}F'^{i}(t),
\label{eqn4}
\end{equation}
and $\chi '^{i}(t)=N^{1/2}(t)\chi^i (t)$, $F'^{i}(t)=NF^i
+\frac{1}{2}(\bar{\psi}\chi^i -\psi\bar{\chi}^i)$.

Integrating over the Grassmann variables and making the following 
redefinition of the odd fields 
$\lambda \rightarrow\frac{1}{3}R^{-1/2}\lambda$ and 
$\chi^i\rightarrow R^{-3/2}\chi^i$, we find the Lagrangian 

\begin{eqnarray}
L&=&-\frac{3R(DR)^2}{\kappa^2N}+\frac{2}{3}i\bar{\lambda}D\lambda 
+\frac{\sqrt{k}}{\kappa}R^{1/2}(\bar{\psi}\lambda -\psi\bar{\lambda})
+\frac{1}{3}NR^{-1}\sqrt{k}\bar{\lambda}\lambda 
+\frac{3k}{\kappa^2}NR+\frac{R^3(D\phi)^2}{2N}-\nonumber\\
&-&i\bar{\chi}^iD\chi^i -\frac{3}{2}\sqrt{k}NR^{-1}\bar{\chi}^i\chi^i-
-\kappa^2Ng(\phi^i)\bar{\lambda}\lambda
-6\sqrt{k}Ng(\phi^i)R^2 -NR^3V(\phi^i) +\nonumber\\
&+&\frac{3}{2}\kappa^2Ng(\phi^i)\bar{\chi}^i\chi^i
+\frac{i\kappa}{2}D\phi^i(\bar{\lambda}\chi^i+\lambda\bar{\chi}^i)
-2N\frac{\partial^2g(\phi^i)}{\partial\phi^i\phi^j}\bar{\chi}^i\chi^j
-\kappa N\frac{\partial g(\phi^i)}{\partial \phi^i}(\bar{\lambda}\chi^i-
\lambda\bar{\chi}^i) + \nonumber\\
&+&\frac{\kappa}{4}R^{-3/2}(\psi\bar{\lambda}
-\bar{\psi}\lambda)\bar{\chi}^i\chi^i-
\kappa R^{3/2}(\bar{\psi}\lambda -\psi\bar{\lambda})g(\phi^i)
+R^{3/2}\frac{\partial g(\phi^i)}{\phi^i}(\bar{\psi}\chi^i-
\psi\bar{\chi}^i) + \nonumber\\ 
&+& \frac{3\kappa^2N}{8R^3}\chi^i\bar{\chi}^i\chi^j\bar{\chi}^j ,
\label{eqn5}
\end{eqnarray}
(eliminating the  auxiliary fields), where 
$DR=\dot{R}-\frac{i\kappa}{6}R^{-1/2}(\psi\bar{\lambda}+\bar{\psi}\lambda )$ 
and 
$D\phi^i =\dot{\phi^i}-\frac{i}{2}R^{-3/2}(\bar{\psi}\chi^i+\psi\bar{\chi}^i)$
are the supercovariant derivatives, and 
$D\lambda =\dot{\lambda}+\frac{i}{2}V\lambda$, 
$D\chi^i =\dot{\chi^i}+\frac{i}{2}V\chi^i$ are the $U(1)$ covariant 
derivatives. 

In the usual models of hybrid inflation $\phi$ is rolling towards zero. The 
potential is tipically of the form
$$
\rm V(\phi)= V_0 + \frac{1}{2} m^2 \phi^2 + \cdots,
$$
and is dominated by the term $\rm V_0$. When $\phi$ fall below some
critical value $\rm \phi_c$, the other field rolls to its vacuum value
so that $\rm V_0$ disappears and inflation ends, but in our scenario we have 
the opposite case of inverted hybrid inflation, where
$\phi$ roll away from the origin and
the scalar potential for the homogeneous scalar fields, in our scenary become
\begin{eqnarray}
V(\phi^i)&=&2\left(\frac{\partial g(\phi^i)}{\partial \phi^i}\right)^2
-3\kappa^2g^2(\phi^i)\nonumber\\
&=&\frac{1}{2}F_i^2-\frac{3}{\kappa^2R^2}B^2,
\label{eqn6}
\end{eqnarray}
The first of them is the potential for the scalar 
fields in the case of global supersymmetry, the second term is the 
contribution of the local character of supersymmetry, where the bosonic
auxiliary fields $F^i$ and B are
\begin{equation}
F^i=2 \frac{\partial g(\phi^i)}{\partial\phi^i}, \qquad\qquad B=\kappa^2
R g(\varphi^i).
\end{equation}
The potential is not positive semi-definite in contrast 
with the standard supersymmetric quantum mechanics. Unlike the standard 
supersymmetric quantum mechanics, this model allows the supersymmetry 
breaking when the vacuum energy is equal to zero.

The selection rules for the ocurrence of spontaneous supersymmetry breaking are
\begin{eqnarray}
\frac{\partial V(\phi^i)}{\partial\phi^i}&=&4\left[
\frac{\partial g(\phi^i)}{\partial\phi^j}
\left(\frac{1}{2}\frac{\partial^2 g(\phi^i)}{\partial\phi^i\partial\phi^j}
\right)-\left(\frac{\partial g(\phi^i)}{\partial\phi^i}
\right)\left(\frac{3\kappa^2}{2}g(\phi^i)
\right)\right]=0, \quad at ~\phi^i= \phi^i_0 \, ,\nonumber\\
V(\phi^i_0)&=&0\Rightarrow\left[\left(\frac{\partial 
g(\phi^i)}{\partial\phi^i}\right)^2-
\frac{3\kappa^2}{2}g^2(\phi^i)\right]=0 \, ,\nonumber\\
F^i&=&2\frac{\partial g(\phi^i)}{\partial\phi^i}\neq 0, \quad at~ 
\phi^i= \phi^i_0 .
\label{eqn7}
\end{eqnarray}

The first condition implies the existence of a minimum in the scalar 
potential; the second condition is the absence of the cosmological constant, 
and the third condition is for the breaking of supersymmetry. From the 
Lagrangian (\ref{eqn5})
we can identify $m_{\frac{3}{2}}=\kappa^2 g(\phi^i_0)$ 
as the gravitino mass in the effective supergravity theory. The factor $R$ 
in the kinetic term of the scalar factor $-(3/\kappa^2)R\dot{R}^2$ plays the 
role of a metric tensor in the Lagrangian (it is the metric tensor in the 
minisuperspace generates by this model).
Now, we need to contruct the corresponding supersymmetric quantum mechanics,
from which the quantum Hamiltonian operator emerges, and becomes the  central 
piece in our study. The quantization procedure must take into account the  
nature of the Grassmann variables, antisymmetrized them and write the 
bilinear combinations in the form of the commutators, 
this leads to the following quantum Hamiltonian
\begin{eqnarray}
{\cal H}&=&-\frac{\kappa^2}{12R}\pi_R^2-\frac{3k}{\kappa^2}R-
\frac{1}{6}\sqrt{k}[\bar{\lambda},\lambda]
+\frac{\pi_{\phi^i}^2}{2R^3}
-\frac{i\kappa\pi_{\phi^i}}{4R^3}([\bar{\lambda},\chi^i]+
[\lambda ,\bar{\chi^i}])-\frac{\kappa^2}{16R^3}[\bar{\lambda},\lambda ]
[\bar{\chi^i},\chi^i]+ \nonumber\\
&+&\frac{3\sqrt{k}}{4R}[\bar{\chi^i},\chi^i]
+\frac{\kappa^2}{2}g(\phi^i)[\bar{\lambda},\lambda ]+6\sqrt{k}g(\phi^i)R^2+
R^3V(\phi^i)
-\frac{3\kappa^2}{4}g(\phi^i)[\bar{\chi^i},\chi^i] + \nonumber\\
&+&\frac{1}{2}\frac{\partial^2g(\phi^i)}
{\partial\phi^i\partial\phi^j}[\bar{\chi^i},\chi^j]
+\frac{\kappa}{2}\frac{\partial g(\phi^i)}{\partial\phi^i}
([\bar{\lambda},\chi^i]-[\lambda ,\bar{\chi^i}]) - 
\frac{3\kappa^2}{32R^3}[\bar{\chi}^i,\chi^i][\bar{\chi}^j,\chi^j].
\label{eqn8}
\end{eqnarray}
We are going to use the following matrix representation for the operators 
$\lambda, \bar{\lambda}, \chi^i, \bar{\chi^i}$
\begin{eqnarray}
\lambda &=&\sqrt{\frac{3}{2}}\sigma_-\otimes 1\otimes 1\otimes 1, 
\qquad\qquad
\chi^1=\sigma_3\otimes\sigma_-\otimes 1\otimes 1,\nonumber\\
\chi^2&=&\sigma_3\otimes\sigma_3\otimes\sigma_-\otimes 1, \qquad\qquad
\chi^3=\sigma_3\otimes\sigma_3\otimes\sigma_3\otimes\sigma_-,\nonumber\\
\bar{\lambda}&=&-\sqrt{\frac{3}{2}}\sigma_+\otimes 1\otimes 1\otimes 1,
\qquad\qquad
\bar{\chi^1}=\sigma_3\otimes\sigma_+\otimes 1\otimes 1,\nonumber\\
\bar{\chi^2}&=&\sigma_3\otimes\sigma_3\otimes\sigma_+\otimes 1,
\qquad\qquad
\bar{\chi^3}=\sigma_3\otimes\sigma_3\otimes\sigma_3\otimes\sigma_+,
\label{eqn9}
\end{eqnarray}
where $\sigma_{\pm}=\frac{1}{2}(\sigma_1\pm i\sigma_2)$ and the commutators 
involucred  in the quantum Hamiltonian can be written as
\begin{eqnarray*}
[\bar{\lambda},\lambda ]&=&-\frac{3}{2}\sigma_3\otimes 1\otimes 1\otimes 1,
\qquad\qquad
[\bar{\chi^1},\chi^1] = 1\otimes\sigma_3\otimes 1\otimes 1, \\
\left[ \bar{\chi^2},\chi^2\right] &=& 1\otimes 1\otimes\sigma_3\otimes 1,
\qquad\qquad
[\bar{\chi^3},\chi^3 ]=1\otimes 1\otimes 1\otimes\sigma_3.
\end{eqnarray*}

\section{WKB type solutions}

The usual hybrid inflation models  are based on a superpotential of the type 
\begin{equation}
g_{inf}=\lambda (M^2-\psi^2)\phi,
\label{eqn10}
\end{equation}
to this superpotential we add a supersymmetry breaking part \cite{19}

\begin{equation}
g_{SB}=m_{3/2}M_p^2(1+\sqrt{\frac{3}{2}}\frac{\varphi}{M_p}
+\frac{3}{4}\frac{\varphi^2}{M_p^2})
\label{eqn11}
\end{equation}
The ``total superpotential" $g$ is  the sum of these two contributions

\begin{equation}
g(\phi^i)=\lambda\left( M^2-\psi^2\right)\phi+m_{\frac{3}{2}}M_p^2
\left(1+\sqrt{\frac{3}{2}}\frac{\varphi}{M_p}+\frac{3}{4}\frac{\varphi^2}{M_p^2}\right).
\label{12}
\end{equation}
and then the scalar potential 
$V(\phi_i)$ ($\phi_1 =\phi ,\phi_2 =\psi ,\phi_3 =\varphi$) 
can be calculated with the help of the relation (\ref{eqn6})
\begin{eqnarray}
V(\phi ,\psi ,\varphi )&=& 2\lambda^2\left( M^2-\psi^2\right)^2
+8\lambda^2\psi^2\phi^2 + 2m_{3/2}^2M_p^4\left(\sqrt{\frac{3}{2}}\frac{1}{M_p}
+\frac{3}{2} \frac{\varphi}{M_p^2}\right)^2
-\nonumber\\
&-&\frac{3}{M_p^2}\lambda^2\left( M^2-\psi^2\right)^2\phi^2
- 3m_{3/2}^2M_p^2\left( 1+\sqrt{\frac{3}{2}}\frac{\varphi}{M_p}+
\frac{3}{4}\frac{\varphi^2}{M_p^2}\right)^2
- \nonumber\\
&-& 6m_{3/2}\lambda\phi\left( M^2-\psi^2\right)
\left( 1+\sqrt{\frac{3}{2}}\frac{\varphi}{M_p}
+ \frac{3}{4}\frac{\varphi^2}{M_p^2}\right).
\label{eqn13}
\end{eqnarray}

This scalar potential possesses two minimums, a global minimum and a local one.
The global minimum is localized in $\phi_0 =0, \psi_0^2 =M^2, \varphi_0 =0$, 
implying that $V(\phi_0,\psi_0,\varphi_0)=0$ and 
$g(\phi_0,\psi_0,\varphi_0)=m_{3/2}M_p^2$. The local minimum is find 
in $\phi *=-\frac{m_{3/2}M_p^2}{2\lambda M^2}, \psi *=0,
 \varphi *=-\sqrt{\frac{2}{3}}M_p$, with these values we have 
$V(\phi *,\psi *,\varphi *)=2\lambda^2M^4$ and the superpotential in 
the local minimum $g(\phi *,\psi *,\varphi *)=0$.

Using the usual representation for the momentum operators
\begin{eqnarray}
\Pi_R&=&-\frac{6}{\kappa^2}R\dot{R},\qquad 
\hat{\Pi}_R=-i\frac{\partial}{\partial R},
\label{eqn14}\\
\Pi_{\phi_i}&=&R^3\dot{\phi}_i, \qquad \hat{\Pi}_{\phi_i}=
-i\frac{\partial}{\partial\phi_i}.
\label{eqn15}
\end{eqnarray}
The corresponding Wheeler-De Witt equation have the form
\begin{eqnarray}
\tilde{{\cal H}}\Psi =R{\cal H}\Psi&=&\left[-\frac{\kappa^2}{12}\pi_R^2-
\frac{3k}{\kappa^2}R^2-\frac{1}{6}\sqrt{k}[\bar{\lambda},\lambda]R
+\frac{\pi_{\phi^i}^2}{2R^2}-\frac{i\kappa\pi_{\phi^i}}{4R^2}
([\bar{\lambda},\chi^i]+ [\lambda ,\bar{\chi^i}]) - \right.\nonumber\\
&-&\left. \frac{\kappa^2}{16R^2}[\bar{\lambda},\lambda ][\bar{\chi^i},\chi^i]+
\frac{3\sqrt{k}}{4}[\bar{\chi^i},\chi^i]
+\frac{\kappa^2R}{2}g(\phi^i)[\bar{\lambda},\lambda ]
+6\sqrt{k}g(\phi^i)R^3 + \right.\nonumber\\
&+&\left. R^4V(\phi^i)-\frac{3\kappa^2}{4}Rg(\phi^i)[\bar{\chi^i},\chi^i]
+\frac{1}{2}\frac{\partial^2g(\phi^i)}
{\partial\phi^i\partial\phi^j}[\bar{\chi^i},\chi^j]R + \right. \nonumber\\
&+&\left.\frac{\kappa}{2}R\frac{\partial g(\phi^i)}{\partial\phi^i}
([\bar{\lambda},\chi^i]-[\lambda ,\bar{\chi^i}])\right]\Psi =0.
\label{eqn16}
\end{eqnarray}
In  the matrix realization to the operators $\lambda, \chi^i$ on the wave
function $\Psi(R,\phi^i)$ (that have 16 components),
the particular Wheeler-De Witt equation that we are going to consider now is 
for the case $R>>l_{pl}$, and $k=0,1$. Under this situation the term of the 
scalar potential in the Wheeler-De Witt
equation is dominant, and we take the scalar potential to be evaluated in 
the local minimum. For this case the superpotential is zero and the 
Wheeler-De Witt equation that governs the quantum behavior is
\footnote{Due to the matrix realization for the operators, there are two
not null 
components for the wave function, i.e. $\Psi_1$ and $\Psi_{16}$,
but both components have the same WDW equation, we only write one of them. }
\begin{equation}
\left[ -\frac{\kappa^2}{12}\hat{\Pi}_R^2+R^4V(\phi *,\psi *,\varphi *)
\right]\Psi_{16}=\left[ -\frac{1}{12M_p^2}\frac{\partial^2}{\partial R^2}+
\frac{1}{2}R^4 F^2_\phi(\phi *,\psi *,\varphi *)\right]\Psi_{16}=0.
\label{eqn17}
\end{equation}
Using the ansatz for the wave function $\Psi =e^{-if}$ and taking the 
WKB approximation 
$\frac{\partial^2 f}{\partial R^2}<<\left(\frac{\partial f}{\partial R}
\right)^2$, the equation (\ref{eqn17}) lead to the following form to the
wave function
\begin{equation}
\Psi =e^{-i\frac{\sqrt{6}}{3}M_pF_{\phi}R^3}.
\label{eqn18}
\end{equation}
In the semiclassical regimen is well known that
\begin{equation}
\Pi_R=\Psi^*\hat{\Pi}_R\Psi\Rightarrow -6M_p^2R\dot{R}=-\sqrt{6}M_pF_{\phi}R^2
\label{eqn19}
\end{equation}
from this, we can obtain the functional form of the scale factor
\begin{equation}
R=C_0 e^{\frac{\sqrt{6}M^2}{3M_p}\lambda t},
\label{20}
\end{equation}
and the Hubble parameter then take the form
\begin{equation}
H=\frac{\sqrt{6}M^2\lambda}{3M_p}.
\label{eqn21}
\end{equation}

Now we are in position for to calculate the density fluctuations.  
The density fluctuations are 
produced by the inflaton field, the other fields not have role here because 
they are fixed to the origin. Then we have
a polinomial  scalar potential in the inflaton field, assuming that
 $\rm V_0$ dominates, 
\begin{equation}
V=2\lambda^2M^4-6m_{3/2}\lambda M^2\phi-\frac{3\lambda^2M^4}{M_p^2}\phi^2,
\label{eqn22}
\end{equation}

The wave function in the semiclassical region is given by (\ref{eqn18}), 
with $V=\frac{1}{2} F_{\phi}^2$. With the help of the relation 
\begin{equation}
\rm \Pi_{\phi}=R^3\dot{\phi}=\Psi^*\hat{\Pi_{\phi}}\Psi,
\label{phi}
\end{equation}
 it is possible to 
obtain  $\dot{\phi}$ 
\begin{equation}
\rm \dot \phi =-\frac{\sqrt{6}}{3}M_p \frac{\partial F_{\phi}}{\partial \phi}
=-\frac{1}{\sqrt{3}} M_p V^{-\frac{1}{2}} 
\frac{\partial V}{\partial \phi}, 
\end{equation}
explicitly
\begin{equation}
\rm \dot \phi =\sqrt{\frac{3}{2}}m_{3/2}M_p.
\label{eqn23}
\end{equation}
In this way, we can obtain the following expresion for the density fluctuations

\begin{equation}
\frac{\delta\rho}{\rho}\simeq \frac{H^2}{\dot{\phi}}=
\frac{2\sqrt{6}\lambda^2 M^4}{9 m_{3/2}M_p^3}.
\label{eqn24}
\end{equation}
which depend on the gravitino and Plank masses values.

On the other hand, using the global minimum in the potential
$V(\phi_0,\varphi_0,\psi_0)=0$, the eigenstates of the Hamiltonian
(\ref{eqn16}) have sixteen components in the matrix representation that we 
have chosen at the end of the previous section. Using the matrix representation for 
$\lambda,\bar \lambda, \chi$ and $\bar \chi$, one find that $\rm \Psi_{16}$
can have the right behaviour when $R\rightarrow \infty$, being the
following partial differential equation in both cases to $k =0,1$.
\begin{equation}
\left( -\frac{1}{12M_p^2}\hat{\Pi}_R^2+\frac{9}{4}m_{3/2}R\right)\Psi_{16}=0,
\qquad {\mbox for ~ k=0},
\end{equation}
and
\begin{equation}
\left(-\frac{1}{12M_p^2}\hat{\Pi}_R^2+6m_{3/2}M_p^2R^3\right)\Psi_{16}=0,
\qquad {\mbox for ~k=1}.
\end{equation}
The semiclassical solutions of these equations becomes
\begin{equation}
\Psi_{16}=e^{-i2\sqrt{2}  \sqrt{m_{3/2}}M_p R^{3/2}},\qquad k=0,
\end{equation}
\begin{equation}
\Psi_{16}=e^{\frac{-12i}{5}\sqrt{2m_{3/2}}M_p^2R^{5/2}},\qquad k=1.
\end{equation}
The behavior of the  scale factor corresponding to the situations of a flat 
and a closed universe are, respectively,
\begin{equation}
R\sim t^{2/3}\qquad k=0,
\end{equation}
\begin{equation}
R\sim t^2\qquad k=1.
\end{equation}
Then, we obtain for a flat universe, a scale factor as a dust dominated 
universe.
The last case corresponds to a scenario like power law inflation \cite{23},
where the scale factor evolves how a power of time. This type of solution is 
perhaps the most prominent example  of exact solution to the full equations of 
motion (not slow roll aproximation); has the extra advantage that 
the equations 
for the generation of density perturbations also can be solved exactly.
The inflaton field acts as a perfect fluid with 
$\omega =\frac{1}{3}$, considering the state equation $p=(\omega -1)\rho$.

\section{Conclusions}
The  inverted hybrid inflation process appear in natural way in the 
supersymmetric theory given in \cite{18,19,20}. Under this scheme, was 
possible to find the behaviour for the scale factor R. When the
scalar potential is evaluated in the local minimum, the behaviour was 
inflationary and the density fluctuations  depend of the inverse of the 
gravitino mass, see  Eq. (\ref{eqn24}). Also, we find that for the global 
minimun in the scalar potential, we obtain for a flat universe, a scale 
factor as a dust dominated universe; and for a closed
universe, the behaviour correspond to a scenario like power law inflation,
where the scale factor evolves how a power of time and the inflaton field acts 
as a perfect fluid. 

Our solutions for
the scale factor are very robust, in the sense that even keeping the next
contribution to the WDW equation (cuadratic terms in R, for instance in 
$\rm k=1$ case) the functional form of the scale factor is retain.

On the other hand, an interesting situation is obtained as consequence of the 
contribution of the fermionic sector. It is possible a tunneling wave function
for flat universe. The wave equation describing the proccess is
$$
\rm \left[ {\hat \Pi}^2_R - \left( \frac{6}{\kappa^2} + \frac{27}{8} \right)
\frac{1}{R^2} + \frac{6 m_{3/2}}{\kappa^2} R- \frac{12 V_0}{\kappa^2} R^4
\right] \Psi_{16} = 0,
$$
for which we can identify the potential
$$
\rm V(R) = - \left( \frac{6}{\kappa^2} + \frac{27}{8} \right)
\frac{1}{R^2} + \frac{6 m_{3/2}}{\kappa^2} R- \frac{12 V_0}{\kappa^2} R^4 .
$$
A schematic form of this potential is show in Fig. 1. 
\begin{figure}[ht]
\centerline{\psfig{file=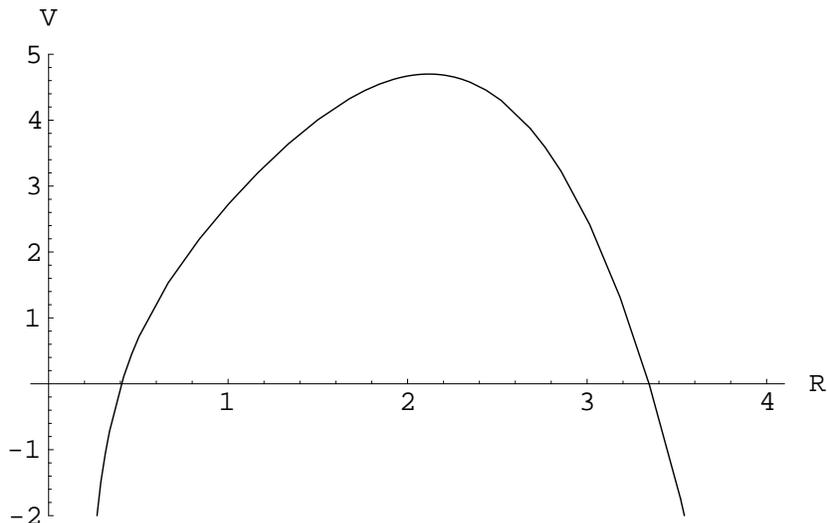,width=11.cm} }
\caption{The structure of the potential 
$\rm V(R) = - 0.08 R^4 -0.2 R^{-2} + 3 R$, with the coefficient of $\rm R^4$
more small than the other terms.}
\end{figure}

\bigskip
\noindent {\bf Acknowledgments}\\
We  thank L. Ure\~na  for critical reading of the manuscript. This work 
was partially supported by CONACyT grant 37851,
PROMEP and Gto. University projects.

\end{document}